\documentclass{epl}

\title{Geometrical Resonance in Spatiotemporal Systems}
\author{J. A. Gonz\'{a}lez\inst{1} \and A. Bellor\'{\i}n\inst{2} \and L. I. Reyes\inst{3} \and C. V\'{a}squez\inst{1} \and L. E. Guerrero\inst{3}}
\institute{
  \inst{1} Centro de F\'{\i }sica, Instituto Venezolano de Investigaciones Cient\'{\i }ficas, Apartado Postal 21827, Caracas 1020-A, Venezuela\\
      \inst{2} Escuela de F\'{\i}sica, Facultad de Ciencias, Universidad Central de Venezuela, Apartado Postal 47586, Caracas 1041-A, Venezuela\\
            \inst{3} Departamento de F\'{\i }sica, Universidad Sim\'{o}n Bol\'{\i}var, Apartado Postal 89000, Caracas 1080-A, Venezuela
	    }

\begin{document}

\maketitle

\begin{abstract}
\noindent We generalize the concept of geometrical resonance to perturbed
sine-Gordon, Nonlinear Schr\"{o}dinger and Complex Ginzburg-Landau
equations. Using this theory we can control different dynamical patterns.
For instance, we can stabilize breathers and oscillatory patterns of large
amplitudes successfully avoiding chaos. On the other hand, this method can
be used to suppress spatiotemporal chaos and turbulence in systems where
these phenomena are already present. This method can be generalized to even
more general spatiotemporal systems.
\end{abstract}

Spatiotemporal chaos \cite{Kaneko,Petrov,Chate,Willeboordse,Gonzalez2} \ is
one of the most important (and \ most studied) phenomena of recent years.
Chaos can be advantageous in some situations, while in many
other situations, it should be avoided or controlled \cite%
{Auerbach,Ott,Auerbach2,Gang,Ditto,Qu,Battogtokh,Grigoriev,Gonzalez}. In
certain cases, the desired effect is a high-amplitude periodic oscillation.
We should drive a nonlinear system with a large external force to produce
such a high-amplitude oscillation. However, this should be done in such a
way that chaos is avoided. Different feedback mechanisms have been devised
to control chaos \cite{Ott,Meyer,Garfinkel,Schiff}. A great deal of research
has been dedicated also to the problem of suppresing chaos by harmonic (or
just periodic) perturbations \cite%
{Qu,Gonzalez,Vohra,Braiman,Azevedo,Chacon,Ding,Chacon2,Chacon4,Chacon5}%
. Among those works are the ones that use the concept of Geometrical
Resonance 
(GR) \cite{Gonzalez,Chacon,Chacon2,Chacon4,Chacon5,Chacon6,Tereshko}%
.

In Ref. \cite{Gonzalez} the concept of GR was used as a chaos-eliminating
mechanism for the perturbed $\varphi^{4}$ equation. In this letter we
generalize the concept of Geometrical Resonance to a very general class of
spatiotemporal systems which includes the sine-Gordon, Nonlinear Schr\"{o}%
dinger, Boussinesq, Toda lattice and Complex Ginzburg-Landau equations
(among others). We will use this concept as a \ method of chaos control when
these equations are nonintegrable because of the presence of perturbations.
GR is an
extension of the linear notion of resonance to a nonlinear formulation based
on a local energy conservation requirement \cite{Chacon2}.

Let us consider the partial differential equation 
\begin{equation}
K_{0}\left[ \phi\right] +K_{1}\left[ \phi,x\right] =q\left( x,t\right) P%
\left[ \phi\right] ,  \label{I}
\end{equation}
where $K_{0}\left[ \phi\right] $ and $P\left[ \phi\right] $ are functions of 
$\phi$, and its derivatives: $\phi_{t}$, $\phi_{x}$, $\phi_{tt}$, $\phi_{xx}$%
, etc. 
Equation 
$K_{0}\left[ \phi\right] =0$
is an integrable Hamiltonian system.
This can be, for instance, the sine-Gordon equation (SGE),
the Nonlinear Schr\"{o}dinger equation (NLSE),
the Boussinesq equation,
or the Toda lattice \cite{Kivshar,Scott}.
On the other hand, $K_{1}\left[ \phi,x\right] $ includes dissipative terms
and $q\left( x,t\right) P\left[ \phi\right] $ is a very general driving
force \cite{Gonzalez2,Gonzalez3,Gonzalez4,Gonzalez5}.

At GR the amplitude, frequency, and space-time shape of $q\left(
x,t\right) $ must satisfy some conditions so that some dynamical properties
of the conservative system are preserved. We will call $\phi_{GR}(x,t)$ a GR
solution of Eq. (\ref{I}) if 
\begin{equation}
K_{1}\left[ \phi_{GR},x\right] =q\left( x,t\right) P\left[ \phi _{GR}\right]
.  \label{III}
\end{equation}

This implies a local energy conservation requirement. The energy integral
that is conserved for equation $K_{0}\left[ \phi\right] =0$
is locally conserved for Eq. (\ref{I})
if condition (\ref{III}) holds. We can use this condition as a mechanism for
chaos control when an additional condition holds: the GR solution must be an
asymptotically stable solution of the (full) equation (\ref{I}). This
condition is introduced here for the first time. We will call Eq. (\ref{III}%
) the exact GR condition and the solutions that satisfy this condition
will be called GR solutions.

We can consider the energy of the system as a ``local almost adiabatic
invariant'' \cite{Arnold}. Then we can write an approximate 
GR condition 
\begin{equation}
\left\langle \frac{dH}{dt} \right\rangle _{T^{\prime}} \simeq0,
\label{IV}
\end{equation}
where $H$ is the energy of the system and $T^{\prime}$ is the period of the
chosen solution of equation $K_{0}\left[ \phi\right] =0$.

As an example, let us investigate the well-known driven and damped SGE 
\begin{equation}
\phi_{tt}-\phi_{xx}+\gamma\phi_{t}+\sin\phi=q(x,t).  \label{VII}
\end{equation}

Suppose the task is to produce breathers of large amplitudes without
entering a chaotic regime.
The exact breather solution to the unperturbed SGE is
\begin{equation}
\phi\left( x,t\right) =4\arctan\left[ \frac{\sqrt{1-\omega^{2}}\sin\left(
\omega t\right) }{\omega\cosh\left( \sqrt{1-\omega^{2}}x\right) }\right] ,
\label{VIII}
\end{equation}
where $\omega$ is arbitrary in the interval $\omega^{2}<1$.

The external force $q(x,t)$ satisfies the GR condition (\ref{III}) when 
\begin{equation}
q\left( x,t\right) =q_{GR}\left( x,t\right) \equiv \frac{4\gamma\sqrt{1-\omega^{2}}\cos\left( \omega
t\right) }{\cosh\left( \sqrt{1-\omega^{2}}x\right) +\left( \frac{1-\omega
^{2}}{\omega^{2}}\right) \left[ \frac{\sin^{2}\left( \omega t\right) }{\cosh%
\left( \sqrt{1-\omega^{2}}x\right) }\right] }.  \label{IX}
\end{equation}

In Eq. (\ref{VII}) if $q(x,t)$ is given by (\ref{IX}), the function (\ref%
{VIII}) is an exact solution of the complete Eq. (\ref{VII}). When the
parameters that define the perturbation (\ref{IX}) are fixed, there is only
one frequency for which function (\ref{VIII}) is the solution. This
frequency is determined by that appearing in (\ref{IX}).

\begin{figure}[t]
\centerline{\includegraphics[width=2in,angle=-90]{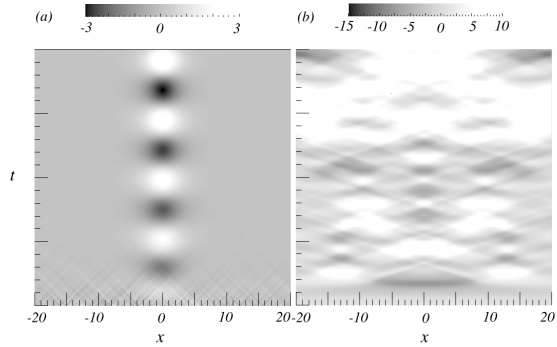}}
\caption{Robustness of the breather in Eq. (\ref{VII}). (a) Even with %
random initial conditions the breather is %
reorganized ($\omega = 0.707$, $\gamma = 0.45$). %
In the numerical simulations with the discretized equation %
the initial conditions were produced by a pseudorandom number %
generator of uniformly distributed values %
in the interval %
$\left[ -1,1\right] $. %
(b) Irregular dynamics produced with a perturbation where the amplitude %
$A$, the frequency $\omega$ and the range $Q$ are not close to %
satisfy the GR condition ($A = 4.5$, $\omega = 0.707$, $Q = 6.6$).}
\label{fig1}
\end{figure}

It is not difficult to show that the solution (\ref{VIII}) is asymptotically
stable in the framework of the full Eq. (\ref{VII}) with $q(x,t)$ 
given by Eq. (\ref{IX}).
In the framework of the unperturbed SGE breathers form a continuum of
solutions similar to the periodic solutions around the fixed points called
centers in Dynamical Systems theory. These solutions are stable in the sense
of Lyapunov but they are not asymptotically stable.
However, the breather solution (\ref{VIII}) in the framework of Eq. (\ref%
{VII}) with $q(x,t)$ given by Eq. (\ref{IX}) is a spatiotemporal limit cycle.
That is, this is a spatiotemporal attractor. All close initial conditions
(in all space configurations) for $t\rightarrow \infty $ will tend to behave
as this solution. This phenomenon is shown in Fig. \ref{fig1}(a).

Perturbation $q_{GR}\left( x,t\right) $ 
can be approximated by a function
of type 
$q(x,t)=f(t)g(x)$ where $g(x)$ is a bell-shaped function and $f(t)$ is a
time-periodic function. This kind of perturbations has been used in several
studies of the SGE \cite{Bishop,Eilbeck}.

The general study of Eq. (\ref{VII}) using the GR concept and the breather
solutions leads to the following conclusions: 
We can avoid chaos with amplitudes A of $q(x,t)$ for which 
$\left| A\right| \leq 4\gamma\sqrt{1-\omega^{2}}$
and $\omega^{2}<1$.
On the other hand,
the range $Q$ of the function $g(x)$ (i.e. the interval of $x$ where $g(x)$
is not exponentially small) should be $Q \leq \frac{1}{\sqrt{1-\omega^{2}}}
$.

\begin{figure}[t]
\centerline{\includegraphics[width=3in]{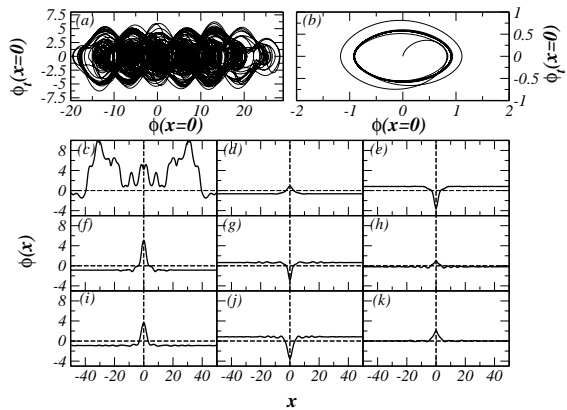}}
\caption{Suppression of spatiotemporal chaos using a given chaos-suppressing
excitation in Eq. (\ref{X}) $\left( f_{0}=0.91,\omega _{d}=0.6\right) $. (a)
Well-developed spatiotemporal chaos when $f_{c}=0$. (b) Controlled dynamics
with $f_{c}=0.4$, $\omega _{c}=0.6$, $\theta =\pi /2$.
Controlling spatiotemporal chaos using a localized excitation in Eq. (%
\ref{XI}): (c) Spatial profile for a given time moment corresponding to
spatiotemporal chaos when  $F_{c}\left( x,t\right) =0$. (d)-(p) Spatial
profiles for different time instants when $F_{c}\left( x,t\right) $ is given
by Eq. (\ref{IX}), $\gamma =0.1,f_{0}=0.5,\omega _{d}=0.6,\omega =0.6$.
The time instants are (d) $t = 0.97$, (e) $t = 7.9$, (f) $t = 15.4$,
(g) $t = 21$, (h) $t = 26.2$, (i) $t = 36.8$, (j) $t = 42$, (k) $t = 46.8$.}
\label{fig2}
\end{figure}

In some cases, when these conditions are not satisfied, the breather
is not stabilized and we can get an irregular behavior. Figure \ref{fig1}(b)
shows an example of the dynamics produced by Eq. (\ref{VII}) with
$q\left( x,t\right) =A\cos \left( \omega t\right) \left[ \cosh \frac{x}{Q}%
+\left( \frac{1-\omega ^{2}}{\omega ^{2}}\right) \left( \frac{\sin
^{2}\left( \omega t\right) }{\cosh \frac{x}{Q}}\right) \right] ^{-1}$
where $A = 4.5$, $\omega = 0.707$, $Q = 6.6$.

There is a wealth of works \cite{Bishop,Eilbeck} dedicated to the numerical
investigation of perturbed sine-Gordon equations using external forces of
type $q(x,t)=f(t)g(x)$. All the results are in agreement with our
theoretical \ results.

Sometimes we have the task of suppressing chaos using a given harmonic
perturbation. Consider e.g. the following equation:
\begin{equation}
\phi_{tt}-\phi_{xx}+\gamma\phi_{t}+\sin\phi=f_{0}\sin\left( \omega
_{d}t\right) +f_{c}\sin\left[ \omega_{c}t+\theta\right] ,  \label{X}
\end{equation}
where $f_{0}\sin\left( \omega_{d}t\right) $ is a chaos-producing excitation,
while $f_{c}\sin\left[ \omega_{c}t+\theta\right] $ is a chaos-suppressing
excitation. The parameters of the chaos-suppressing excitation should be
determined.
In this case we can use the condition (\ref{IV}) to find
parameters for the chaos-suppressing perturbation. Fig. \ref{fig2} (a)-(b)
shows an
example of chaos control using this technique. 

We would like to stress here that the force (\ref{IX}) can be used to
control a well-developed spatiotemporal chaos.

Let us consider the following equation:
\begin{equation}
\phi_{tt}-\phi_{xx}+\gamma\phi_{t}+\sin\phi=f_{0}\sin\left( \omega
_{d}t\right) +F_{c}\left( x,t\right) .  \label{XI}
\end{equation}

When $F_{c}\left( x,t\right) \equiv 0$, the system 
presents spatiotemporal chaos for
$-\infty <x<\infty $, see Fig. \ref{fig2}(c). Now, if we turn on the controlling force $F_{c}\left(
x,t\right) $ defined as function (\ref{IX}), we obtain a very regular
spatiotemporal pattern as that shown in Fig. \ref{fig2} (d)-(k).
The most important remark here is that we are controlling spatiotemporal
chaos in the whole space using a localized perturbation.
We should add here that other works have used localized perturbations to
control spatiotemporal chaos \cite{Kocarev,Boccaletti,Wu}.

The damped and ac-driven NLSE 
\begin{equation}
i\phi_{t}+\phi_{xx}+2\left| \phi\right| ^{2}\phi+i\alpha \phi=\varepsilon
e^{i\omega t}  \label{XIII}
\end{equation}
is another fundamental model in many areas of physics \cite{Kaup,Bishop2,Cai}%
.
At sufficiently large $\varepsilon$ the dynamics of this model becomes
chaotic \cite{Cai}.

Suppose we have a general driving term: 
$i\phi_{t}+\phi_{xx}+2\left| \phi\right| ^{2}\phi+i\alpha \phi=q\left(
x,t\right)$ . 
We will take the one-soliton solution of unperturbed NLSE \cite{Scott} as a GR
solution: Then $q(x,t)$ must satisfy the condition:

\begin{equation}
q_{GR}\left( x,t\right) =\frac{\alpha\sqrt{\omega}e^{i\left( \omega
t+\pi/2\right) }}{\cosh\left( \sqrt{\omega}x\right) }.  \label{XV}
\end{equation}
where $\omega$ can be any positive number.
Thus, if the perturbation is localized and the amplitude $\varepsilon$
satisfies the condition $\varepsilon\approx\alpha\sqrt{\omega}$, then the
chaotic regime can be avoided. The one-soliton solution of NLSE is
stabilized.
We should remark that this can be achieved also by other localized
perturbations.

If we use the two-soliton breather solution as a GR solution, we can obtain
another driving force satisfying a GR condition: 
\begin{equation}
q_{GR}\left( x,t\right) =\frac{4\alpha\left[ \cosh\left( 3x\right)
+3e^{i8t}\cosh\left( x\right) \right] e^{i\left( t+\pi/2\right) }}{%
\cosh\left( 4x\right) +4\cosh\left( 2x\right) +3\cos\left( 8t\right) }.
\label{XVI}
\end{equation}

See Ref. \cite{Scott} for a discussion of multisoliton solutions. As in
Eq. (\ref{XV})
we can introduce here an arbitrary parameter $\omega $. However, in this
case, we are more interested in the relationship between the two intrinsic
frequencies of the solution.
This force can be approximated by a function of type 
\begin{equation}
F\left( x,t\right) =\varepsilon_{1}g_{1}(x)e^{i\left(
\omega_{1}t+\pi/2\right) }+\varepsilon_{2}g_{2}(x)e^{i\left(
\omega_{2}t+\pi/2\right) },  \label{XVII}
\end{equation}
where $\omega_{1}=1$, $\omega_{2}=9$, and $g_{1}(x)$ and $g_{2}(x)$ are
localized functions.

In Ref. \cite{Cai} a breather was stabilized using a two-frequency drive: 
$F\left( x,t\right) =\varepsilon _{1}e^{i\omega _{1}t}+\varepsilon%
_{2}e^{i\omega _{2}t}$,
where $\omega _{1}=1$ and $\omega _{2}=9$. This result can be seen as a
confirmation of the GR approach for the NLSE. We should add that the
phenomenon of breather stabilization is quite robust. For instance, if $%
\omega _{1}=1$, in addition to $\omega _{2}=9$, other close frequencies can
be used, namely $\omega _{2}=8$ and $\omega _{2}=10$. This means that the GR
condition can be satisfied approximately and that the Eq. (\ref{IV})
can also be used as a guide for the search of a controlling force.
As in the case of the breather of the SGE, the breather of the
NLSE is asymptotically stable. We have checked numerically that the
nonchaotic breather solution is the most stable one. Unfortunately we have
no space to show a picture.

Regarding localized excitations we should emphasize that GR analysis
explains diverse fundamental results on stability of localized solutions
previously obtained by perturbation theory \cite{Gonzalez,Gonzalez2,Cai}
including those relative
to one-soliton and two-soliton solutions of the NLSE. In this sense, in
future works, it would be interesting to consider the case of the N-soliton
solutions (see e.g. Appendix B, Ref. \cite{Scott}).

The control of spatiotemporal chaos (or turbulence) in the Complex
Ginzburg-Landau equation (CGLE) \cite%
{Battogtokh,Kuramoto,Hohenberg,Chate2,Aranson, Xiao,Kramer} is a problem of
great practical interest \cite{Kramer}.
We are interested in the modified CGLE \cite{Battogtokh,Hohenberg}: 
\begin{equation}
\phi_{t}=\phi+\left( 1+ic_{1}\right) \phi_{xx}-\left( 1-ic_{3}\right) \left|
\phi\right| ^{2}\phi+F_{c}\left( x,t\right) .  \label{XIX}
\end{equation}

The term $F_{c}\left( x,t\right) $ is the control signal. Without the
control signal $\left( F_{c}\left( x,t\right) =0\right) $, the turbulence
develops when the Benjamin-Feir condition $1-c_{1}c_{3}<0$ is satisfied.
This equation can be rewritten in the following form: 
\begin{equation}
i\phi_{t}+c_{1}\phi_{xx}+c_{3}\left| \phi\right| ^{2}\phi=i\left( \phi
_{xx}+\phi-\left| \phi\right| ^{2}\phi\right) +iF_{c}\left( x,t\right) .
\label{XX}
\end{equation}
When the right hand side of Eq. (\ref{XX}) is zero, it reduces to the
NLSE.

If $\phi\left( x,t\right) =f\left( x\right) \exp\left( i\omega t\right) $
is a soliton solution of the NLSE, then we can use the following controlling
signal 
\begin{equation}
F_{c}\left( x,t\right) =\left[ f^{3}\left( x\right) -f\left( x\right)
-f_{xx}\left( x\right) \right] \exp\left( i\omega t\right) .  \label{XXI}
\end{equation}

Equation (\ref{XIX}) (with $F_{c} \equiv 0$) presents turbulence for
$c_{1}=2$, $c_{3}=0.8$. We have been able to suppress this turbulence
using the $F_{c}\left( x,t\right)$ given by Eq. (\ref{XXI}) with
$\omega =12$ and $f\left( x\right)$ is the one-soliton solution of
equation $c_{1}f_{xx}-\omega f+c_{3}f^{3}=0$ \cite{Scott}.

In this context, we should explain that in some cases, the stabilization
process can require a force that is not a small perturbation. Furthermore,
this technique can be used both as a way to stabilize a
pre-existing solution of the unperturbed system and as a way to impose an
arbitrary solution to the system. However, the success of all these
endeavors depends on a very important fact: the final solution should be an
asymptotical stable solution of the perturbed system. Incidentally, we
should mention that the stabilization of unstable plane waves in the CGLE
can be done using a nonlinear diffusion term \cite{Montagne}.

In some situations we can apply some perturbations using technological means
in order to satisfy the stability conditions. Nevertheless, we should say
that, very often, nature itself can apply the controlling perturbations.
There are many natural regimes described by the mentioned equations
in the presence of perturbations where the resulting dynamics is not
chaotic. Our results can provide an explanation for these phenomena.

Numerous observations and experiments show that elastic waves from natural
phenomena and human-made machines may alter water and oil production 
\cite{Beresnev}%
. In some cases wave excitation may appreciably increase the mobility of
these fluids. A new technology \cite{Beresnev} based on these experiments is used
to stimulate the \ reservoir as a whole. Here seismic frequency waves are
applied at the earth's surface by arrays of vibrators. Many of the phenomena
involved in this effect are described by the equations discussed in this
paper, namely: NLSE, SGE, Boussinesq equation and other equations of type (%
\ref{I}) (see Refs. \cite{Nikolaevskiy}). For the optimization of the method, it is
necessary to sustain spatiotemporal nonlinear oscillations of the reservoir
with some frequency and shape. Based on ideas related to the results
presented in this paper we have designed a new technology using 
a specific geometrical arrangement of the surface vibrators %
\cite{Gonzalez6}. 

The nonlinear PDE
possess an infinite number of different solutions. Among them one can choose
a feasible one in order to implement. Even if only a given type of
perturbation is allowed due to technical limitations, it is
always possible to use the approximate condition (\ref{IV}) as in the case of
task (\ref{X}).

The concept that links all the situations where we have been able to
suppress chaos is based on the mutual cancellation of nonintegrable terms as
described by equations (\ref{III}) and (\ref{IV}). In other words, we should
add some temporal perturbation in such a way that (at least approximately)
both the dissipative and the total driving terms mutually cancel. A
remarkable
situation (which is a particular case of the general theory but, at the same
time, is present in all the studied systems) is that of breather-like
oscillations. These patterns can be stabilized using some spatially
localized time-periodic perturbations, where the amplitude, the spatial
range and the frequency must satisfy some relationship. However, this
phenomenon is robust. A fine-tuning is not necessary. There is always a
whole valid interval of values for the amplitude, range and frequency that
produces qualitatively the same result.

The most common
perturbation in scientific research is $F\left( t\right) =f_{0}\cos\left(
\omega t\right) $. However, nature is very rich in dynamical behaviors.
Our work shows that using very general spatiotemporal perturbations $F\left(
x,t\right) $ we can make the difference between regular or chaotic behavior.
Using certain spatiotemporal perturbations $F\left( x,t\right) $ we can
stabilize a breather or we can produce a turbulent dynamics. We have been
able to control different patterns in the sine-Gordon, Nonlinear Schr\"{o}%
dinger, and Complex Ginzburg-Landau equations. 
Each of these systems
possesses wide applications in many areas of Physics. Furthermore we believe
that these ideas can be applied to other systems.

A. Bellor\'{\i}n would like to thank CDCH-UCV for support under 
project PI-03-11-4647-2000.


\begin{thebibliography}{0}

\bibitem{Kaneko}
  \Name{Kaneko K.}
  \REVIEW{Phys. Rev. Lett.}{63}{1989}{219}.

\bibitem{Petrov}
  \Name{Petrov V. et al.}
  \REVIEW{Phys. Rev. Lett.}{75}{1995}{2895}.

\bibitem{Chate}
  \Name{Chat\'{e} H.}
  \REVIEW{Physica D}{86}{1995}{238}

\bibitem{Willeboordse}
  \Name{Willeboordse F. H. \and Kaneko K.}
   \REVIEW{Physica D}{86}{1995}{428}

\bibitem{Gonzalez2}
  \Name{Gonz\'{a}lez J. A., Guerrero L. E. \and Bellor\'{\i}n A.}
  \REVIEW{Phys. Rev. E}{54}{1996}{1265}

\bibitem{Auerbach} 
  \Name{Auerbach D. et al.}
  \REVIEW{Phys. Rev. Lett.}{69}{1992}{3479}

\bibitem{Ott} 
  \Name{Ott E., Grebogi C. \and Yorke J. A.}
  \REVIEW{Phys. Rev. Lett.}{64}{1990}{1196}

\bibitem{Auerbach2}
  \Name{Auerbach D.}
  \REVIEW{Phys. Rev. Lett.}{72}{1994}{1184}

\bibitem{Gang} 
  \Name{Gang H. \and Zhilin Q.}
  \REVIEW{Phys. Rev. Lett.}{72}{1994}{68}

\bibitem{Ditto} 
  \Name{Ditto W. L., Spano M. L. \and Lindner J. F.}
  \REVIEW{Physica D}{86}{1995}{198}

\bibitem{Qu}
  \Name{Qu Z. et al.}
  \REVIEW{Phys. Rev. Lett.}{74}{1995}{1736}

\bibitem{Battogtokh} 
  \Name{Battogtokh D. \and Mikhailov A.}
  \REVIEW{Physica D}{90}{1996}{84}

\bibitem{Grigoriev} 
  \Name{Grigoriev R. O., Cross M. C. \and Schuster H. G.}
  \REVIEW{Phys. Rev. Lett.}{79}{1997}{2795}

\bibitem{Gonzalez} 
  \Name{Gonz\'{a}lez J. A. et al.}
  \REVIEW{Phys. Rev. Lett.}{80}{1998}{1361}

\bibitem{Meyer} 
  \Name{Meyer J. R., Kruer M. R. \and Bartoli F. J.}
  \REVIEW{J. Appl. Phys.}{51}{1980}{5513}

\bibitem{Garfinkel}
  \Name{Garfinkel A., Spano M. L. \and Ditto W. L.}
  \REVIEW{Science}{257}{1992}{1230}

\bibitem{Schiff} 
  \Name{Schiff S. J., Jerger K., \and Duong D. H.}
  \REVIEW{Nature}{370}{1994}{615}

\bibitem{Vohra} 
  \Name{Vohra . T., Fabiny  L. \and Bucholtz F.}
  \REVIEW{Phys. Rev. Lett.}{75}{1995}{65}

\bibitem{Braiman}
  \Name{Braiman Y. \and Goldhirsch I.}
  \REVIEW{Phys. Rev. Lett.}{66}{1991}{2545}

\bibitem{Azevedo} 
  \Name{Azevedo A. \and Rezende S. M.}
  \REVIEW{Phys. Rev. Lett.}{66}{1991}{1342}
    
\bibitem{Chacon} 
  \Name{Chac\'{o}n R. \and D\'{\i}az Bejarano J.}
  \REVIEW{Phys. Rev. Lett.}{71}{1993}{3103}

\bibitem{Ding} 
  \Name{Ding W. X. et al.}
  \REVIEW{Phys. Rev. Lett.}{72}{1994}{96}

\bibitem{Chacon2} 
  \Name{Chac\'{o}n R.}
  \REVIEW{Phys. Rev. Lett.}{77}{1996}{482}

\bibitem{Chacon4} 
  \Name{Chac\'{o}n R.}
  \REVIEW{J. Math. Phys.}{38}{1997}{1477}

\bibitem{Chacon5} 
  \Name{Chac\'{o}n R.}
  \REVIEW{Phys. Rev. Lett.}{86}{2001}{1737}

\bibitem{Chacon6}
  \Name{Chac\'{o}n R.}
  \REVIEW{Phys. Rev. E}{54}{1996}{6153}

\bibitem{Tereshko}
  \Name{Tereshko V. \and Shchekinova E.}
  \REVIEW{Phys. Rev. E}{58}{1998}{423}

\bibitem{Kivshar} 
  \Name{Kivshar Y. S. \and Malomed B. A.}
  \REVIEW{Rev. Mod. Phys.}{61}{1989}{763}

\bibitem{Scott} 
  \Name{Scott A.}
  \Book{Nonlinear Science}
  \Publ{Oxford University Press, New York}
  \Year{1999}

\bibitem{Gonzalez3} 
  \Name{Gonz\'{a}lez J. A. \and Ho\l yst J. A.}
  \REVIEW{Phys. Rev. B}{45}{1992}{410338}

\bibitem{Gonzalez4} 
  \Name{Gonz\'{a}lez J. A. \and Mello B. A.}
  \REVIEW{Phys. Lett. A}{219}{1996}{226}

\bibitem{Gonzalez5} 
  \Name{Gonz\'{a}lez J. A., Bellor\'{\i}n A. \and Guerrero L. E.}
  \REVIEW{Phys. Rev. E}{65}{2002}{065601(R)}

\bibitem{Arnold} 
  \Name{Arnold V. I.}
  \Book{Geometrical Methods in the Theory of Ordinary Differential Equations}
  \Publ{Springer-Verlag, New York}
  \Year{1989}

\bibitem{Bishop} 
  \Name{Bishop A. R. et al.}
  \REVIEW{Phys. Rev. Lett.}{50}{1983}{1095}

\bibitem{Eilbeck} 
  \Name{Eilbeck J. C., Lomdahl P. S. \and Newell A. C.}
  \REVIEW{Phys. Lett. A}{87}{1981}{1}

\bibitem{Kocarev}
  \Name{Kocarev L., Tasev Z. \and Parlitz U.}
  \REVIEW{Phys. Rev. Lett.}{79}{1997}{51}

\bibitem{Boccaletti}
  \Name{Boccaletti S., Bragard J. \and Arecchi F. T.}
  \REVIEW{Phys. Rev. E}{59}{1999}{6574}

\bibitem{Wu}
  \Name{Wu S., He K. \and Huang Z.}
  \REVIEW{Phys. Lett. A}{260}{1999}{345}

\bibitem{Kaup} 
  \Name{Kaup D. J. \and Newell A. C.}
  \REVIEW{Phys. Rev. B}{18}{1978}{5162}

\bibitem{Bishop2} 
  \Name{Bishop A. R. et al.}
  \REVIEW{SIAM J. Math. Anal.}{21}{1990}{1511}

\bibitem{Cai} 
  \Name{Cai D. et al.}
  \REVIEW{Phys. Rev. E}{49}{1994}{R1000}

\bibitem{Kuramoto} 
  \Name{Kuramoto Y.}
  \Book{Chemical Oscillations, Waves and Turbulence}
  \Publ{Springer-Verlag, New York}
  \Year{1984}

\bibitem{Hohenberg} 
  \Name{Cross M. C. \and Hohenberg P. C.}
  \REVIEW{Rev. Mod. Phys.}{65}{1993}{851}

\bibitem{Chate2} 
  \Name{Chat\'{e} H. \and Manneville P.}
  \REVIEW{Physica A}{224}{1996}{348}

\bibitem{Aranson} 
  \Name{Aranson I., Levine H. \and Tsimring L.}
  \REVIEW{Phys. Rev. Lett.}{72}{1994}{2561}

\bibitem{Xiao} 
  \Name{Xiao J. et al.}
  \REVIEW{Phys. Rev. Lett.}{81}{1998}{5552}

\bibitem{Kramer} 
  \Name{Aranson I. S. \and Kramer L.}
  \REVIEW{Rev. Mod. Phys.}{74}{2002}{99}

\bibitem{Montagne}
  \Name{Montagne R. \and Colet P.}
  \REVIEW{Phys. Rev. E}{56}{1997}{4017}

\bibitem{Beresnev}
  \Name{Beresnev I. A. \and Johnson P. A.}
  \REVIEW{Geophysics}{59}{1994}{1000}, and references therein

\bibitem{Nikolaevskiy}
  \Name{Nikolaevskiy V. N.}
  \Book{Geomechanics and Hydrodynamics with Applications to %
Reservoir Engineering}
  \Publ{Kluwer Academic Publisher, Boston}
  \Year{1996}, and references therein

\bibitem{Gonzalez6}
  \Name{Gonz\'{a}lez J. A. et al.}
  \REVIEW{to be published}{}{2003}{}

\end{thebibliography}
\end{document}